\documentstyle[prb,preprint,aps]{revtex}
\begin{document}
\draft
\title{Effect of interchain coupling on conducting polymer
luminescence: excimers in derivatives of poly(phenylene vinylene)}
\author{M. W. Wu and E. M. Conwell}
\address{Center for Photoinduced Charge Transfer, Chemistry Department,
University of Rochester, Rochester, New York 14627\\
and Xerox Corporation, Wilson Center for Technology, 114-22D, Webster,
New York 14580$^*$}
\maketitle

\begin{abstract}
Optical excitation of a chain in a polymer film may result in
formation of an excimer, a superposition of on-chain excitons
and charge-transfer excitons on the originally excited chain and a
neighboring chain. The excimer emission is red-shifted compared to
that of an on-chain exciton by an amount depending on the interchain coupling
$t_\perp$. Setting up the excimer wavefunction and calculating the
red shift, we determine average $t_\perp$ values, referred to a
monomer, of 0.52\ eV and 0.16\ eV for
poly(2,5-hexyloxy $p$-phenylene cyanovinylene), CN-PPV, and
poly[2-methoxy, 5-(2$^\prime$-ethyl-hexyloxy)-1, 4 $p$-phenylene
vinylene], MEH-PPV, respectively, and use them to
determine the effect of interchain distance on the emission.
\end{abstract}


The existence of polaron pairs or charge transfer excitons in
poly($p$-phenylene vinylene), PPV, and derivatives was first
suggested to explain the magnetic field
 effect on photoconductivity\cite{fran}
and the long-lived photoinduced absorption found in PPV and some
of its derivatives.\cite{hsu} Wavefunctions and energy levels of charge
transfer (CT) excitons have been calculated
for PPV.\cite{miz1,con1,con2} As noted
in Ref.\ \onlinecite{miz1}, CT excitons would show little absorption
or emission polarized parallel to the chains because there is
little component of the dipole moment along the chains. The
observation of strong absorption or emission polarized parallel to
the chains\cite{hsu} requires that the wavefunction
be a superposition of exciton wavefunctions along the
two chains with the CT exciton wavefunctions,\cite{miz1}
 {\em i.e.}, an excimer.\cite{6}
The mixture of these wavefunctions results in a red shift of the
emission of the excimer compared to that of an isolated chain.\cite{6}
This shift can be determined by comparing emission from a film
with that from a dilute solution of the polymer. Such a red shift
has been seen in CN-PPV,\cite{sam} ladder poly(paraphenylene),\cite{pau}
(L-PPP) and a series of $\pi$-conjugated polybenzobisthiazoles.\cite{jen}

Photoinduced absorption measurements on MEH-PPV films,\cite{yan} and L-PPP
films,\cite{pau} revealed a photogenerated species with a long lifetime,
$\sim$ ns, at least several times as long as the radiative lifetime.
The suggestion was made that the photogenerated species
is polaron pairs;\cite{hsu} it was reinforced by the observation that
these long-lived pairs are not generated in dilute
solution.\cite{yan,kra,pau}
Conwell has maintained that the long-lived pairs are excimers
both in the cases of MEH-PPV and CN-PPV.\cite{con,conwell}
 A calculation of the
most probable configurations of adjacent chains of MEH-PPV,\cite{con} and
of CN-PPV,\cite{con} showed that emission is much
less probable from excimers
of the former polymer because of the large interchain distance.
Recently, Samuel {\em et al.} reported that the emission from MEH-PPV
is also downshifted in going from solution to film\cite{sam97}
although the shift is
less than half as much as that seen for CN-PPV. Also, they find that
the emissive state is longer lived in film than in solution.\cite{sam97}
 However, the
emission from MEH-PPV retains some vibronic structure, which is not
expected for an excimer and indeed not seen for CN-PPV. They suggest,
therefore, that the emission from MEH-PPV is also due to an
interchain excitation, but the interchain interaction is
weaker than that in CN-PPV. This is just what should be expected in view
of the larger interchain distance 4.1\ \AA~found for MEH-PPV as compared
to 3.4\ \AA~for CN-PPV.\cite{con}

In the much studied case of excimers formed by molecules, such as
pyrene, in solution the molecules assume the intermolecular distance
corresponding to the minimum energy for the excimer, determined
by the balance of attractive $\pi$-wavefunction overlap and core
repulsion. In a polymer film constraints on the chains result in a range
of interchain distances and relative orientations, none of them
necessarily optimum for excimer formation. The result is, understandably,
a wide range of excimer emission frequencies and efficiencies, with
the possibility of retention of the phonon structure for weakly coupled
chains.

To calculate the frequency of the excimer emission from PPV derivatives,
we use a simplified microscopic model for a PPV chain introduced by
Rice and Gartstein.\cite{rice} The two chains are taken to be finite but
very long, parallel and coterminous. Chain deformation is neglected.
The excimer wave function is taken to be the sum of exciton
wavefunctions on chain 1 and chain 2, $|\Psi_1\rangle$ and
$|\Psi_2\rangle$ respectively, and CT excitons $|\Psi_{12}\rangle$
with the hole on chain 1, the electron on chain 2, and $|\Psi_{21}\rangle$
with the position of electron and hole reversed. Setting up these
wavefunctions, in what follows we determine the coefficient $R$
of $|\Psi_{12}\rangle$ and $|\Psi_{21}\rangle$ in this sum by a variational
method. The value of $R$ depends on the interchain coupling. Comparing our
predicted red shift with the experimental shift, we determine average
values of $t_\perp$ for CN-PPV and MEH-PPV. The former is found to be
several times as large, as is appropriate to the smaller
interchain distance for that case.

We start from a  model Hamiltonian for the two coupled
conjugated polymer chains: $H=H_1+H_2+V_{12}
+H_\perp$. $H_i$ ($i=1$, 2) describes intrachain interactions
for the $i$th chain. $V_{12}$ stands for the long-range Coulomb
interaction between electron (e) and hole (h), separated on
different chains. $H_\perp$ represents the coupling of the two
chains. The $H_i$ are given by\cite{rice}
\begin{eqnarray}
H_i&=&\sum_{n\sigma}[\alpha_0(a_{in\sigma}^\dagger a_{in\sigma}+
b_{in\sigma}^\dagger b_{in\sigma})-t(a_{in+1\sigma}^\dagger a_{in\sigma}
+b_{in+1\sigma}^\dagger b_{in\sigma}
+\mbox{H.C.})]\nonumber\\
\label{hi}
&&\mbox{}-\sum_{mn\sigma s}U(n-m)b_{in\sigma}^\dagger
b_{in\sigma}a_{ims}^\dagger a_{ims}\;,
\end{eqnarray}
where $b_{in\sigma}^\dagger$ and $a_{in\sigma}^\dagger$
create an electron in the conduction band and a
hole in the valence band, respectively, each with spin $\sigma$,
located at the $n$th monomer in the $i$th chain.
In writing this Hamiltonian, we measure energies from the center of
the gap, so $\alpha_0$ denotes the distance from midgap of the center
of each band. $t$ is the hopping integral between
nearest neighbour monomers. From the first term of Eq.\ (\ref{hi})
one can get the single
electron and single hole energy spectra: $\varepsilon_{ike}=\varepsilon
_{ikh}=\alpha_0-2t\cos(ka)$, with $a$ representing the length of a
monomer. Accordingly, $W=4t$ is the band width. From the excitation
spectra, neglecting the Coulomb interaction, one can easily get the
energy gaps for e-h excitation, {\em ie.}, the minimum energy
required to  create an e-h pair on one polymer chain, $E_g=2
(\alpha_0-2t)$. The quantity
\begin{equation}
U(n-m)=U\delta_{nm}+\frac{V}{|n-m|}(1-\delta_{nm})
\end{equation}
represents the intrachain Coulomb interaction, with $U$ standing for the
on-monomer e-h interaction and $V$ the nearest-neighbour monomer e-h
interaction. The interchain Coulomb interaction is given by
\begin{equation}
V_{12}=-\sum_{mn\sigma s}V_1(n-m)
b_{2n\sigma}^\dagger b_{2n\sigma}a_{1ms}^\dagger a_{1ms}
-\sum_{mn\sigma s}V_1(n-m)
b_{1n\sigma}^\dagger b_{1n\sigma}a_{2ms}^\dagger a_{2ms}\;,
\end{equation}
with
\begin{equation}
V_{1}(n-m)=\frac{V_1}{\sqrt{(n-m)^2+B}}\;.
\end{equation}
Here $B=\kappa_\Vert d^2/(\kappa_\perp a^2)$ and $d$
stands for the perpendicular distance between 
the two chains.\cite{landau}
$\kappa_\Vert$ and $\kappa_\perp$ are dielectric constants along
and perpendicular to the polymer chains, respectively.
The Hamiltonian
\begin{equation}
\label{hperp}
H_\perp=t_\perp\sum_{n\sigma}(b_{1n\sigma}^\dagger b_{2n\sigma}+
\mbox{H.C.})+t_\perp
\sum_{n\sigma}(a_{1n\sigma}^\dagger a_{2n\sigma}+\mbox{H.C.})
\end{equation}
represents the single-particle interchain
hopping process. We assume that $t_\perp$ falls off exponentially
according to\cite{miz}
\begin{equation}
\label{tp}
t_\perp=t_0 e^{-\mu d}=t_0 \exp (-\mu a\sqrt{B
\kappa_\perp/\kappa_\Vert})\;.
\end{equation}

The intrachain wavefunction $|\Psi_i\rangle$ ($i=1$, 2) for the $i$th
chain can be acquired as follows. We first construct a two-particle
wave function of the singlet e-h pair excitation on the $i$th
chain as\cite{rice}
\begin{equation}
|\Psi_i\rangle=\frac{1}{\sqrt{2}}\sum_{mn\sigma}\Phi_{nm}a_{in\sigma}
^\dagger b_{im-\sigma}^\dagger|0\rangle
\end{equation}
with $|0\rangle$ denoting the ground state. $\phi_{nm}$ is
assumed to be real and is normalized according to $\sum_{nm}\Phi_{nm}^2
=1$. From $H_i|\Psi_i\rangle=E|\Psi_i\rangle$, we obtain the
lowest excitonic bound state with energy $E_1$,\cite{rice}
\begin{equation}
\label{eq1}
(E-2\alpha_0)\Phi_{n-m}=-2t(\Phi_{n-m+1}+\Phi_{n-m-1})
-U(n-m)\Phi_{n-m}\;.
\end{equation}
Similarly, we can construct two-particle charge-transfer (CT)
wavefunctions as
\begin{equation}
\label{p12}
|\Psi_{12}\rangle=\frac{1}{\sqrt{2}}\sum_{nm\sigma}\phi_{nm}
a_{1n\sigma}^\dagger b_{2m-\sigma}^\dagger |0\rangle
\end{equation}
and
\begin{equation}
|\Psi_{21}\rangle=\frac{1}{\sqrt{2}}\sum_{nm\sigma}\phi_{nm}^\prime
a_{2n\sigma}^\dagger b_{1m-\sigma}^\dagger |0\rangle\;.
\end{equation}
$\phi_{nm}$ and $\phi_{nm}^\prime$ can differ from each other
by a global phase. Here we take this phase as 1 so $\phi_{nm}=\phi_{nm}
^\prime$. $\phi_{nm}$ can be determined from $(H_1+H_2+V_{12})|\Psi_{12}
\rangle=E|\Psi_{12}\rangle$ and is given by the equation
\begin{equation}
\label{eqct}
(E-2\alpha_0)\phi_{n-m}=
-2t(\phi_{n-m+1}+\phi_{n-m-1})-V_1(n-m)\phi_{n-m}\;,
\end{equation}
with the normalization condition $\sum_{n}{\phi
_{n}}^2=1/N$ (Here we have assumed $\phi_{n}$ to be real).
Again we only need the lowest CT excitonic bound state with energy $E_2$
from Eq.\ (\ref{eqct}).

The variational wave function of the excimer $|\Psi\rangle$ can
be constructed from the wavefunctions of the on-chain exciton
$|\Psi_i\rangle$ with exciton energy $E_1$ and the
wave functions of CT exciton $|\Psi_{12}\rangle$ and $|\Psi_{21}\rangle$
with CT exciton energy $E_2$:
\begin{equation}
\label{psi}
|\Psi\rangle=\frac{1}{\sqrt{2+2R^2}}(|\Psi_1\rangle+a|\Psi_2\rangle
+R|\Psi_{12}\rangle+R^\prime |\Psi_{21}\rangle)\;.
\end{equation}
Here $|a|=1$ and $|R|=|R^\prime|$. We further assume all quantities
are real. It can be seen
that only if $R=R^\prime$ and $a=1$ can
it be possible to have the total energy $E=\langle\Psi|H|\Psi
\rangle<E_1$. Then
\begin{equation}
\label{energy}
E=(E_1+R^2E_2+2\Gamma R)/(1+R^2)\;,
\end{equation}
where $\Gamma\equiv\langle\Psi_i|H_\perp|\Psi_{12}\rangle=
\langle\Psi_i|H_\perp|\Psi_{21}\rangle$.
$R$ is determined by minimizing $E$, which gives
\begin{equation}
\label{r}
R=-\frac{\sqrt{(E_2-E_1)^2+4\Gamma^2}-(E_2-E_1)}{2\Gamma}\;.
\end{equation}

For the numerical calculations, we employ a chain of $N=400$ unit
cells for each of the polymers and use periodic boundary conditions.
The values of various unknown parameters must be chosen to give the
correct gap and exciton binding energy. For
MEH-PPV the optical absorption edge is at 2.1\ eV.\cite{yu1}
The single particle energy gap of MEH-PPV has been measured as
2.45\ eV.\cite{cam} The exciton binding energy is the difference
of these two numbers,\cite{conwell1} thus 0.35\ eV. The intrachain
Coulomb potential coefficient $V$ is of the order of
the Coulomb attraction between an electron and a hole separated
by one monomer, which is $\sim 1/2$\ eV.\cite{rice}
Choosing $\alpha_0=3.2$\ eV, $t=U=1$\ eV,\cite{rice}
and $V=0.44$\ eV, we calculate for MEH-PPV
$E_g=2.4$\ eV, the exciton creation energy
$E_1=2.065$\ eV and its binding energy $\epsilon_b=0.335$\ eV.
For CN-PPV the optical absorption edge is at $\sim
2.3$\ eV.\cite{sam} It is reasonable to assume that the
exciton binding energy is similar to that of PPV and MEH-PPV. With
$\alpha_0=3.3$\ eV, $t= U=1$\ eV,
and $V=0.4$\ eV, we obtain $E_g=2.6$\ eV, the exciton
creation energy $E_1=2.285$\ eV and
$\epsilon_b=0.315$\ eV. To evaluate the CT wave function we use
$\kappa_\Vert=8$, $\kappa_\perp=3$ and
$V_1=V$. It is also necessary to have a value for $t_\perp$
to obtain the CT wavefunctions $|\Psi_{12}\rangle$ and $|\Psi_{21}\rangle$
and the CT exciton energy from Eqs.\ (\ref{p12}) through (\ref{eqct}).
In principle it is possible to calculate a value for $t_\perp$ given
the positions of all the atoms in a pair of monomers on adjacent chains.
We could use the atom positions determined by the procedure of
Ref.\ \onlinecite{con} to give minimum energy. The resulting $t_\perp$
would not be representive of the film, however, because its amorphous
nature implies, as discussed earlier, a wide range of interchain
distances and relative orientations. We therefore determined what could be
considered an average $t_\perp$ by finding what value is required
to obtain from the equations and parameters given above the red shift
of the excimer emission. The result, referred to a monomer in each
case, was $t_\perp=0.52$\ eV for CN-PPV films, 0.16\ eV for MEH-PPV films.

To determine the dependence on $d$ of the excimer emission it is
necessary to know how $t_\perp$ decreases with increasing distances
between the chains. For PPV the coupling between C atoms on
neighboring chain is mediated by H atoms, the C-H distances
being the smallest. Also the wavefunction of the $2p$ level on H, which
is coupled to the carbon $p$ orbital,\cite{go} decays much less rapidly
than the carbon $p$ orbital. The resulting value of $\mu$ in
Eq.\ (\ref{tp}) is 1.18\ \AA$^{-1}$. \cite{miz} The change of exciton
emission energy $E$ with interchain distance for this
value of $\mu$ is shown in Fig.\ 1. It is seen that as $d$ increases, $E$
tends rapidly to the exciton energy $E_1$. We plot also in Fig.\ 1
the probable contribution of the CT exciton to the excimer state,
$P=R^2/(1+R^2)$, as a function of distance $d$.
It is seen that, at the most probable interchain
distance for CN-PPV, $P\sim 1/2$, but decreases rapidly with
increasing $d$. For MEH-PPV $P$ is always smaller. If direct
coupling between C atoms on neighboring chains dominates over C-H
coupling, $\mu$ would be 2.013\ \AA$^{-1}$, \cite{con} and the changes
with distance would be even more rapid.

In summary, we have set up wave functions for excimers in CN-PPV
and MEH-PPV, and compared them with experimental red shifts
of the excimer emission to evaluate the average interchain
transfer integral in films of these materials. Due to the large
electronegativity of the CN group, which makes the interchain
distance smaller, $t_\perp$ is much larger for that case.
We use the $t_\perp$ values to determine how the emission frequency
varies with interchain spacing.

We acknowledge the support
of the National Science Foundation
under Science and Technology Center grant CHE912001.

\begin{figure}
\caption{Plots of the emission energy $E$ (solid curves) and probability
of CT exciton contribution to the excimer wave function, $P$ (dot-dashed
 curves)
for CN-PPV and MEH-PPV. Dotted lines show the exciton energy $E_1$.}
\end{figure}


\begin{references}

\bibitem[*]{byline}Mailing address.
\bibitem{fran} E.L. Frankevich {\em et al.}, Phys. Rev. B {\bf 46},
9320 (1992).
\bibitem{hsu} J.W.P. Hsu {\em et al.}, Phys. Rev. B {\bf 49}, 712 (1994).
\bibitem{miz1} H.A. Mizes and E.M. Conwell, Phys. Rev. B {\bf 50},
11243 (1994).
\bibitem{con1} E.M. Conwell and H.A. Mizes, Phys. Rev. B {\bf 51}, 6953
(1995).
\bibitem{con2} E.M. Conwell and H.A. Mizes, Synth. Met. {\bf 69}, 613 (1995).
\bibitem{6} See, {\em eg.}, I. Michl and V. Bonacic-Koutecky, {\em
Electronic Aspects of Organic Photochemistry}, (Wiley, New York, 1990).
\bibitem{sam} I.D. Samuel, G. Rumbles, and C.J. Collison, Phys. Rev. B
{\bf 52}, R115n
73 (1995).
\bibitem{pau} T. Pauck {\em et al.}, Chem. Phys. Lett. {\bf 244}, 171
(1995).
\bibitem{jen} S.A. Jenekhe and J.A. Osaheni, Science {\bf
265}, 765 (1994).
\bibitem{yan} M. Yan {\em et al.}, Phys. Rev. Lett. {\bf 72}, 1104
(1994).
\bibitem{kra} B. Kraabel {\em et al.}, Mol. Cryst. Liq.
Cryst. {\bf 256}, 733 (1994).
\bibitem{con} E.M. Conwell, J. Perlstein, and S. Shaik, Phys. Rev. B
{\bf 54}, R2308 (1996).
\bibitem{conwell} E.M. Conwell, Synth. Met. {\bf 84}, 497 (1997).
\bibitem{sam97} I.D.W. Samuel {\em et al.}, Synth. Met. {\bf 84},
497 (1997).
\bibitem{rice} M.J. Rice and Yu.N. Gartstein, Phys. Rev. B {\bf 53},
10764 (1996).
\bibitem{landau} L.D. Landau and E.M. Lifshitz, {\em Electrodynamics of
Continuous Media} (Addison-Wesley, Reading, MA 1960), p.\ 62.
\bibitem{miz} H.A. Mizes and E.M. Conwell, Mol. Cryst. Liq. Cryst.
{\bf 256}, 519 (1994).
\bibitem{yu1}  G. Yu, C. Zhang, and A.J. Heeger, Appl. Phys. Lett.
{\bf 64}, 1540 (1994).
\bibitem{cam} I.H. Campbell, T.W. Hagler, D.L. Smith, and
J.P. Ferraris, Phys. Rev. Lett. {\bf 76}, 1900 (1996).
\bibitem{conwell1} E.M. Conwell, Synth. Met. {\bf 83}, 101 (1996).
\bibitem{go} P.Gomes da Costa, R.G. Dandrea, and E.M. Conwell,
Phys. Rev. B {\bf 47},
1800 (1993).

\end{references}
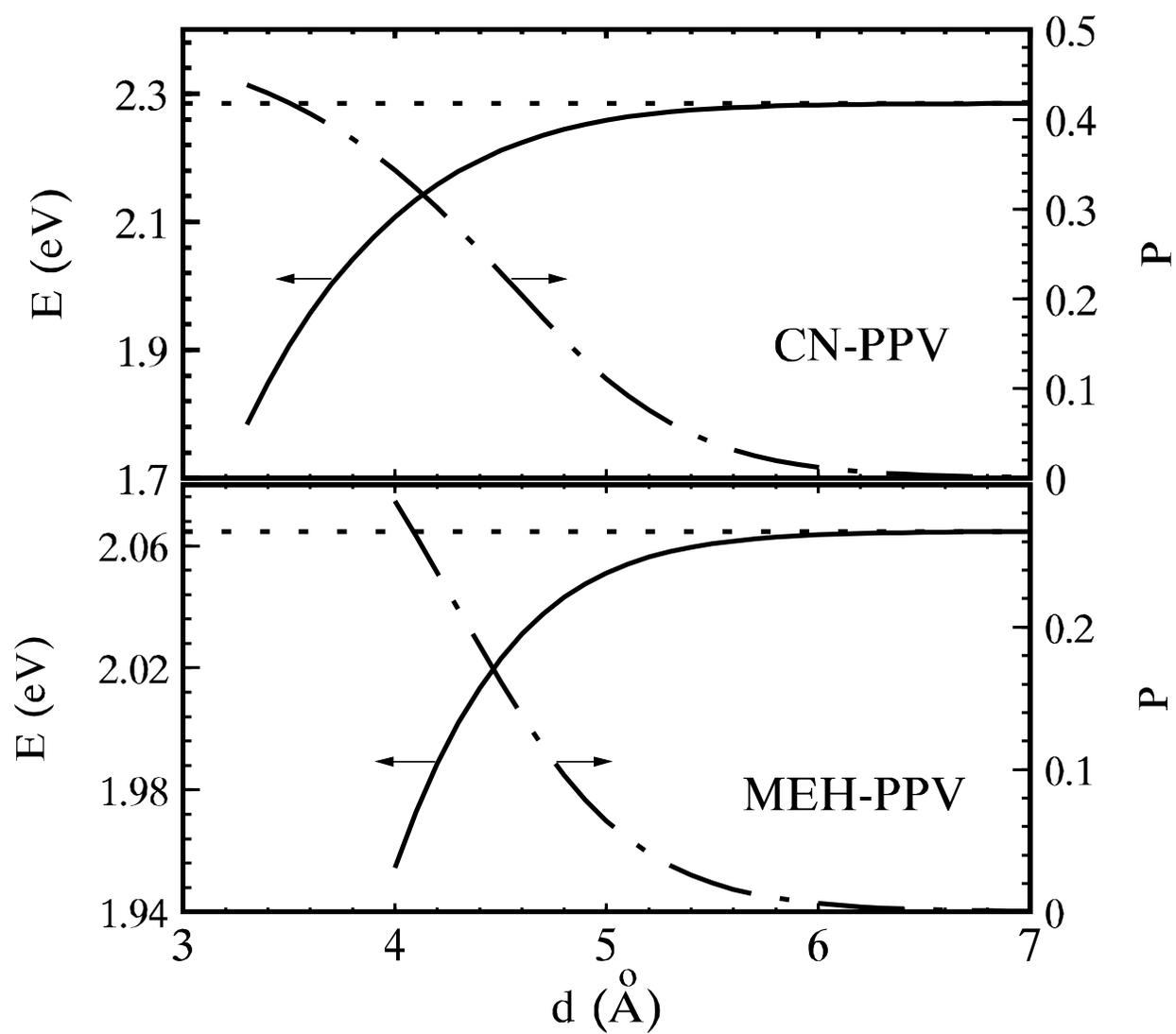
\end{document}